\documentclass[pre,showpacs,preprintnumbers,amsmath,amssymb]{revtex4}

\usepackage{epsfig}
\usepackage{graphicx}
\usepackage{dcolumn}
\usepackage{bm}

\begin{document}

\title{EXPANDING THE TRANSFER ENTROPY TO IDENTIFY INFORMATION CIRCUITS IN COMPLEX SYSTEMS  }

\author{
S. Stramaglia$^{1,2}$, Guo-Rong Wu$^{3,4}$, M. Pellicoro$^{1,2}$,
and D. Marinazzo$^{3}$} \affiliation{$^1$ Istituto Nazionale di
Fisica Nucleare, Sezione di Bari, Italy\\} \affiliation{$^2$
Dipartimento di Fisica, University of Bari,
Italy\\}\affiliation{$^3$ Faculty of Psychology and Educational
Sciences, Department of Data Analysis, Ghent University, Henri
Dunantlaan 1, B-9000 Ghent, Belgium \\}
  \affiliation{$^4$ Key Laboratory for
NeuroInformation of Ministry of Education, School of Life Science
and Technology, University of Electronic Science and Technology of
China, Chengdu 610054, China.\\}
\date{\today}

\begin{abstract}
We propose a formal expansion of the transfer entropy to put in
evidence irreducible sets of variables which provide information for
the future  state of each assigned target. Multiplets characterized
by a large contribution to the expansion are associated to
informational circuits present in the system, with an informational
character which can be associated to the sign of the contribution.
For the sake of computational complexity, we adopt the assumption of
Gaussianity and use the corresponding exact formula for the
conditional mutual information. We report the application of the
proposed methodology on two EEG data sets. 
\end{abstract}

\maketitle
\section{INTRODUCTION}
The inference of couplings between dynamical subsystems, from data,
is a topic of general interest. Transfer entropy \cite{te}, which is
related to the concept of Granger causality \cite{granger}, has been
proposed to distinguish effectively driving and responding elements
and to detect asymmetry in the interaction of subsystems. By
appropriate conditioning of transition probabilities this quantity
has been shown to be superior to the standard time delayed mutual
information, which fails to distinguish information that is actually
exchanged from shared information due to common history and input
signals \cite{lehnertz,wibral}.  On the other hand, Granger
 formalized the notion that, if the prediction of one time
series could be improved by incorporating the knowledge of past
values of a second one, then the latter is said to have a {\it
causal} influence on the former. Initially developed for econometric
applications, Granger causality has gained popularity also in
neuroscience (see, e.g.,
\cite{blinoska,smirnov,dingprl,noiprl,faes}). A discussion about the
practical estimation of information theoretic indexes for signals of
limited length can be found in \cite{porta}. Transfer entropy and
Granger causality are equivalent in the case of Gaussian stochastic
variables \cite{barnett}: they  measure the information flow between
variables \cite{hla}. Recently it has been shown that the presence
of redundant variables influences the estimate of the information
flow from data, and that maximization of the total causality is
connected to the detection of groups of redundant variables
\cite{noired}.

In recent years, information theoretic treatment of groups of
correlated degrees of freedom have been used to reveal their
functional roles as memory structures or those capable of processing
information \cite{borst}. Information theory suggests quantities
that reveal if a group of variables is mutually redundant or
synergetic \cite{sch,bett}. Most approaches for the identification
of functional relations among nodes of a complex networks rely on
the statistics of motifs, subgraphs of {\it k} nodes that appear
more abundantly than expected in randomized networks with the same
number of nodes and degree of connectivity \cite{milo,yeger}.

An interesting approach to identify functional subgraphs in complex
networks, relying on an exact expansion of the mutual information
with a group of variables, has been presented in \cite{bettencourt}.
In this work we generalize these results to show a formal expansion
of the transfer entropy which puts in evidence irreducible sets of
variables which provide information for the future  state of the
target. Multiplets of variables characterized by an high value,
unjustifiable by chance, will be associated to informational
circuits present in the system.
Additionally, in applications where linear models are sufficient to
explain the phenomenology, we propose to use the exact formula for
the conditioned mutual information among Gaussian variables so as to
get a computationally efficient approach. An approximate procedure
is also developed, to find informational circuits of variables
starting from few variables of the multiplet by means of a greedy
search. We illustrate the application of the proposed expansion to a
toy model and two real EEG data sets.

The paper is organized as follows. In the next section we describe
the expansion and motivate our approach. In section III we report
the applications of the approach and describe our greedy search
algorithm. In section IV we draw our conclusions.

\section{Expansion}
We start describing the work in \cite{bettencourt}. Given a
stochastic variable $X$ and a family of stochastic variables
$\{Y_k\}_{k=1}^n$, the following expansion for the mutual
information, analogous to a Taylor series, has been derived there:

\begin{eqnarray}
\begin{array}{l}
S\left(X|\{Y\}\right) - S(X)= -I\left(X;\{Y\}\right)=\\\sum_i
{\Delta S(X)\over \Delta Y_i} + \sum_{i>j} {\Delta^2 S(X)\over
\Delta Y_i \Delta Y_j} + \cdots
+{\Delta^n S(X)\over \Delta Y_i \cdots \Delta Y_n},\\
\end{array}
\label{mi}
\end{eqnarray}

where the variational operators are defined as
\begin{equation}\label{diff1}
{\Delta S(X)\over \Delta Y_i}=S\left(X|Y_i\right) -
S(X)=-I\left(X;Y_i\right),
\end{equation}
\begin{equation}\label{diff2}
{\Delta^2 S(X)\over \Delta Y_i \Delta Y_j}=- {\Delta
I\left(X;Y_i\right)\over \Delta
Y_j}=I\left(X;Y_i\right)-I\left(X;Y_i\right |Y_j),
\end{equation}
\begin{equation}\label{diff3}
{\Delta^3 S(X)\over \Delta Y_i \Delta Y_j \Delta
Y_k}=I\left(X;Y_i|Y_k\right)-I\left(X;Y_i\right |Y_j , Y_k) -
I\left(X;Y_i\right)+I\left(X;Y_i\right |Y_j),
\end{equation}
and so on.

Now, let us consider $n+1$ time series $\{x_\alpha (t)\}_{\alpha
=0,\ldots,n}$. The lagged state vectors are denoted
$$Y_\alpha (t)= \left(x_\alpha (t-m),\ldots,x_\alpha (t-1)\right),$$
$m$ being the window length.

Firstly we may use the expansion (\ref{mi}) to model the statistical
dependencies among the $x$ variables at equal times.  We take $x_0$
as the target time series, and the first terms of the expansion are
\begin{equation}\label{tei}
W_i^0=-I\left(x_0;x_i \right)
\end{equation}
for the first order;
\begin{equation}\label{2orderi}
Z_{ij}^0=I\left(x_0;x_i\right)-I\left(x_0;x_i|x_j\right)
\end{equation}
for the second order; and so on. We note that $$Z_{ij}^0 = - {\cal
I} \left(x_0;x_i;x_j\right),$$ where ${\cal I}
\left(x_0;x_i;x_j\right)$ is the {\it interaction information}, a
well known information measure for sets of three variables
\cite{mcgill}; it expresses the amount of information (redundancy or
synergy) bound up in a set of variables, beyond that which is
present in any subset of those variables. Unlike the mutual
information, the interaction information can be either positive or
negative. Common-cause structures lead to negative interaction
information . As a typical example of positive interaction
information one may consider the three variables of the following
system: the output of an XOR gate with two independent random inputs
(however some difficulties may arise in the interpretation of the
interaction information, see \cite{bell}). It follows that positive
(negative) $Z_{ij}^0$ corresponds to redundancy (synergy) among the
three variables $x_0$, $x_i$ and $x_j$.

In order to go beyond equal time correlations, here we propose to
consider the flow of information from multiplets of variables to a
given target. Accordingly, we consider
\begin{equation}\label{mi2}
S\left(x_0|\{Y_k\}_{k=1}^n\right) -
S(x_0)=-I\left(x_0;\{Y_k\}_{k=1}^n\right),
\end{equation}
which measures to what extent all the remaining variables contribute
to specifying the future state of  $x_0$. This quantity can be
expanded according to (\ref{mi}):
\begin{eqnarray}
\begin{array}{l}
S\left(x_0|\{Y_k\}_{k=1}^n\right) - S(x_0)=\\\sum_i {\Delta
S(x_0)\over \Delta Y_i} + \sum_{i>j} {\Delta^2 S(x_0)\over \Delta
Y_i \Delta Y_j} + \cdots +{\Delta^n S(x_0)\over \Delta Y_i \cdots
\Delta Y_n}.\\
\end{array}
\label{mi3}
\end{eqnarray}


A drawback of the expansion (\ref{mi2}) is that it does not remove
shared information due to common history and input signals;
therefore we choose to condition it on the past of $x_0$, i.e.
$Y_0$. To this aim we introduce the conditioning operator ${\cal
C}_{Y_0}$:
$${\cal C}_{Y_0} S(X) = S(X|Y_0),$$ and observe that ${\cal C}_{Y_0}$ and the variational operators (\ref{diff1}) commute.
It follows that we can condition  the expansion (\ref{mi3}) term by
term, thus obtaining
\begin{eqnarray}
\begin{array}{l}
S\left(x_0|\{Y_k\}_{k=1}^n,Y_0\right) - S(x_0
|Y_0)=-I\left(x_0;\{Y\}_{k=1}^n|Y_0\right)=\\ \sum_i {\Delta
S(x_0|Y_0)\over \Delta Y_i} + \sum_{i>j} {\Delta^2 S(x_0|Y_0)\over
\Delta Y_i \Delta Y_j} + \cdots +{\Delta^n S(x_0|Y_0)\over \Delta
Y_i \cdots \Delta Y_n}.\\
\end{array}
\label{mi4}
\end{eqnarray}


The first order terms in the expansion are given by:
\begin{equation}\label{te}
A_i^0={\Delta S(x_0|Y_0)\over \Delta
Y_i}=-I\left(x_0;Y_i|Y_0\right),
\end{equation}
and coincide with the bivariate transfer entropies $i \to 0$ (times
-1). The second order terms are
\begin{equation}\label{2order}
B_{ij}^0=I\left(x_0;Y_i|Y_0 \right)-I\left(x_0;Y_i|Y_j,Y_0\right),
\end{equation}
and may be seen as a generalization of the interaction information
${\cal I}$; hence a positive (negative) $B_{ij}^0$ corresponds to a
redundant (synergetic) flow of information $\{ i,j\} \to 0$.  The
typical examples of synergy and redundancy, in the present framework
of network analysis, are the same as in the static case, plus a
delay for the flow of information towards the target. The third
order terms are
\begin{eqnarray}
\begin{array}{ll}
C_{ijk}^0=&I\left(x_0;Y_i|Y_j,Y_0 \right)+I\left(x_0;Y_i|Y_k,Y_0
\right)\\&-I\left(x_0;Y_i|Y_0\right)-I\left(x_0;Y_i|Y_j,Y_k,Y_0\right),
\end{array}
\label{3order}
\end{eqnarray}
and so on.

The generic term in the expansion (\ref{mi4}),
\begin{equation}\label{term}
\Omega_k={\Delta^k S(x_0|Y_0)\over \Delta Y_i \cdots \Delta Y_k},
\end{equation}
is symmetrical under permutations of the $Y_i$ and, remarkably,
statistical independence among any of the $Y_i$ results in vanishing
contribution to that order. Therefore each nonvanishing accounts for
an irreducible set of variables providing information for the
specification of the target: the search for for informational
multiplets is thus equivalent to search for terms (\ref{term}) which
are significantly different from zero.  Another property of
(\ref{mi4}) is that the sign of each term is connected to  the
informational character of the corresponding set of variables, see
\cite{bettencourt}).

For practical applications, a reliable estimate of conditional
mutual information from data should be used. Non parametric methods
are recommendable when nonlinear effects are relevant. However, a
conspicuous amount of phenomenology in brain can be explained by
linear models: therefore, for the sake of computational load,  In
this work we adopt the assumption of Gaussianity and use the exact
expression that holds in this case \cite{barnett}, which reads as
follows. Given multivariate Gaussian random variables $X$, $W$ and
$Z$, the conditioned mutual information is
\begin{equation}\label{bar}
I\left(X;W|Z\right)=\frac{1}{2}\ln{|\Sigma(X|Z)|\over|\Sigma(X|W\oplus
Z)|},
\end{equation}
where $|\cdot|$ denotes the determinant, and the partial covariance
matrix is defined
\begin{equation}\label{cov1}
\Sigma(X|Z)=\Sigma(X)-\Sigma(X,Z)\Sigma(Z)^{-1}\Sigma(X,Z)^\top,
\end{equation}
in terms of the covariance matrix $\Sigma(X)$ and the cross
covariance matrix $\Sigma(X,Z)$; the definition of $\Sigma(X|W\oplus
Z)$ is analogous.

The statistical significance of (\ref{term}) can be assessed by
observing that it is the sum of terms like (\ref{bar}) which, under
the null hypothesis $I\left(X;W|Z\right)=0$, have a $\chi ^2$
distribution. Alternatively, statistical testing may be done using
surrogate data obtained by random temporal shuffling of the target
vector $x_0$; the latter strategy is the one we use in this work.

\section{APPLICATIONS}
\subsection{Second order terms}
In this subsection we show the application of the proposed
expansion, truncated at the second order. To this aim  we turn to
real electroencephalogram (EEG) data, the window length $m$ being
fixed by cross validation. Firstly we consider recordings obtained
at rest from 10 healthy subjects. During the experiment, which
lasted for 15 min, the subjects were instructed to relax and keep
their eyes closed. To avoid drowsiness, every minute the subjects
were asked to open their eyes for 5 s. EEG was measured with a
standard 10-20 system consisting of 19 channels \cite{nolte}. Data
were analyzed using the linked mastoids reference, and are available
from \cite{website_nolte}.

For each subject we consider several epochs of 4 seconds in which
the subjects kept their eyes closed. For each epoch we compute the
second order terms at equal times $Z^0_{ij}$ and the lagged ones
$B^0_{ij}$; then we average the results over epochs. In order to
visualize these results, for each target electrode we plot a on a
topographic scalp map the pairs of electrodes which are redundant or
synergetic with respect to it. Both quantities are distributed with
a clear pattern across the scalp. Interactions at equal times are
one order of magnitude higher than the lagged interactions, and are
dominated by the effect of spatial proximity, see fig. 1. On the
other hand, $B^0_{ij}$ show a richer dynamics, such as
interhemispheric communications and predominance redundancy to and
from the occipital channels, see fig. 2, reflecting the prominence
of the occipital rhythms when the subjects rest with their eyes
closed.

As another example we consider intracranial EEG recordings from a
patient with drug-resistant epilepsy and which has thus been
implanted with an array of $8\times 8$ cortical electrodes and two
depth electrodes with six contacts. The data are available at
\cite{kol_web} and described in \cite{kol_paper}. For each seizure
data are recorded from the preictal period, the 10 seconds preceding
the clinical onset of the seizure, and the ictal period, 10 seconds
from the clinical onset of the seizure. We analyze  data
corresponding to eight seizures and average the corresponding
results.

For each electrode we compute the lagged influences $B^0_{ij}$,
obtaining for each electrode the pair of other electrodes with
redundant or synergetic contribution to its future. The patient has
a putative epileptic focus in a deep hippocampal region, with the
seizure that then spreads to the cortical areas. In fig. 3  we
report the values of coefficients $B$ taking as the target a
cortical electrode located on the putative cortical focus: we report
 the values of $B^0_{ij}$ corresponding to
all the couple of the electrodes, as well as their sum over
electrode $j$. It is clear how the redundancy increases during the
seizure. On the other hand, for sensors from 70 to 76, corresponding
to a depth electrode, the redundancy is higher in the preictal
period, reflecting the fact that the seizure is already active in
its primary focus even if not yet clinically observable. The values
of $B$ corresponding to this electrode are reported in fig.  4.

\subsection{Greedy search of multiplets}
Given a target variable, the time required for the exhaustive search
of all the subsets of variables, with a statistically significant
information flow (\ref{term}), is exponential in the size of the
system. It follows that the exact search for large multiplets is
computationally unfeasible, hence we adopt the following approximate
strategy.  We start from
 a pair of variables with non-vanishing second order term $B$ w.r.t. the given target. We consider these two variables as a {\it
 seed}, and aggregate other variables to them so as to construct a multiplet.
The third variable of the subset is selected among the remaining
ones as those that, jointly with the previously chosen variable,
maximize the modulus $|C|$ of the corresponding third order term.
Then, one keeps adding the rest of the variables by iterating this
procedure. Calling $Z_{k-1}$ the selected set of k - 1 variables,
the set $Z_k$ is obtained adding, to $Z_{k-1}$, the variable, among
the remaining ones, with the greatest modulus of $\Omega_k$. These
iterations stop when $\Omega_k$, corresponding to $Z_k$, is not
significantly different from zero (the Bonferroni correction for
multiple comparisons is to be applied at each iteration); $Z_{k-1}$
is then recognized as the multiplet originated by the initial pair
of variables chosen as the seed.

We apply this strategy to the following toy model

\begin{eqnarray}
\begin{array}{lll}
x_0 (t)&=a \;\eta(t-1)+\sigma \xi_0 (t),&\\
x_\alpha (t)&=b_\alpha \; \eta(t)+\sigma_1 \xi_\alpha (t),& \alpha=1,\ldots,m\\
x_\beta (t)&=\sigma_2 \xi_\beta (t), &\beta=m+1,\ldots,m+M\\
\end{array}
\label{map}
\end{eqnarray}

where $\xi$ and $\eta$ are i.i.d. unit variance Gaussian variables.
 In this model the target $x_0$ is
influenced by the process $\eta$; variables $x_\alpha$,
$\alpha=1,\ldots,m$, are a mixture of $\eta$ and noise $\xi$, whilst
the remaining $M$ variables are pure noise. Estimates of $\Omega_k$
are based on time series, generated from (\ref{map}) and 1000
samples long. The results are displayed in figure (\ref{figmodels}).
Firstly we consider the case $m=20$ and $M=0$, with all the twenty
variables driving the target with equal couplings $b_\alpha$; in
figure (\ref{figmodels})-A we depict the term $\Omega_k$
corresponding to the $k$-th iteration of the greedy search. We note
that $\Omega_k$ has alternating sign and its modulus decreases with
$k$. In figure (\ref{figmodels})-B we consider another situation,
with $m=10$ and $M=10$, the ten non-zero couplings $b_\alpha$ being
non-uniform. $\Omega_k$ still shows alternating sign, and $\Omega_k$
vanishes for $k> 9$; hence  the multiplet of ten variables is
correctly identified. The order of selection is related to the
strength of the coupling: variables with stronger coupling are
selected first.

In figure (\ref{figC3}) we consider again the EEG data from healthy
subjects with closed eyes \cite{website_nolte}, and apply the greedy
search taking C3 as the target and $\{C4, C6 \}$ as the seed. We
find a subset of 9 variables influencing the target. The fact that
the sign of $\Omega_k$ is alternating, as in the previous model,
suggests that the channels in this set correspond to a single source
which is responsible for the inter-hemispheric communication towards
the target electrode C3. In figure (\ref{figO1}) we take O1 as the
target and $\{F3, C5 \}$ as the seed. A subset of 11 variables is
found which describes the information flow from the frontal to the
occipital cortex.

\begin{figure}[ht]
\includegraphics[width=8.5cm]{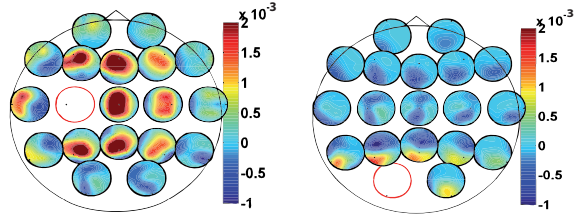}
\caption{{\rm The instantaneous components $Z^0_{ij}$ for two target
electrodes, C3 on the left and O1 on the right. The target electrode
is in white, and for each of the other electrodes $i$ on the map,
the value of $Z^0_{ij}$ is displayed for the other electrodes.
\label{figEEGZ}}}\end{figure}

\begin{figure}[ht]
\includegraphics[width=8.5cm]{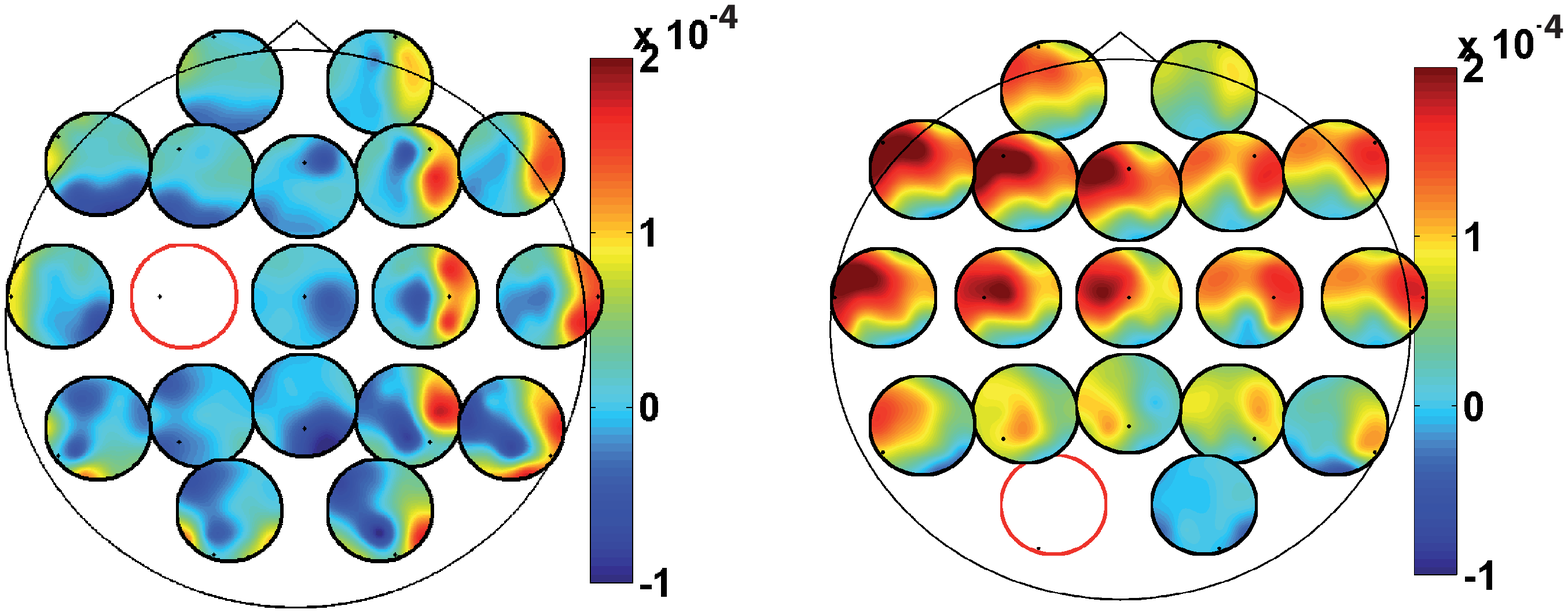}
\caption{{\rm The lagged components $B^0_{ij}$ for two target
electrodes, C3 on the left and O1 on the right. The target electrode
is in white, and for each of the other electrodes $i$ on the map,
the value of $B^0_{ij}$ is displayed for the other electrodes.
\label{figEEGZ2}}}\end{figure}

\begin{figure}[ht]
\includegraphics[width=8.5cm]{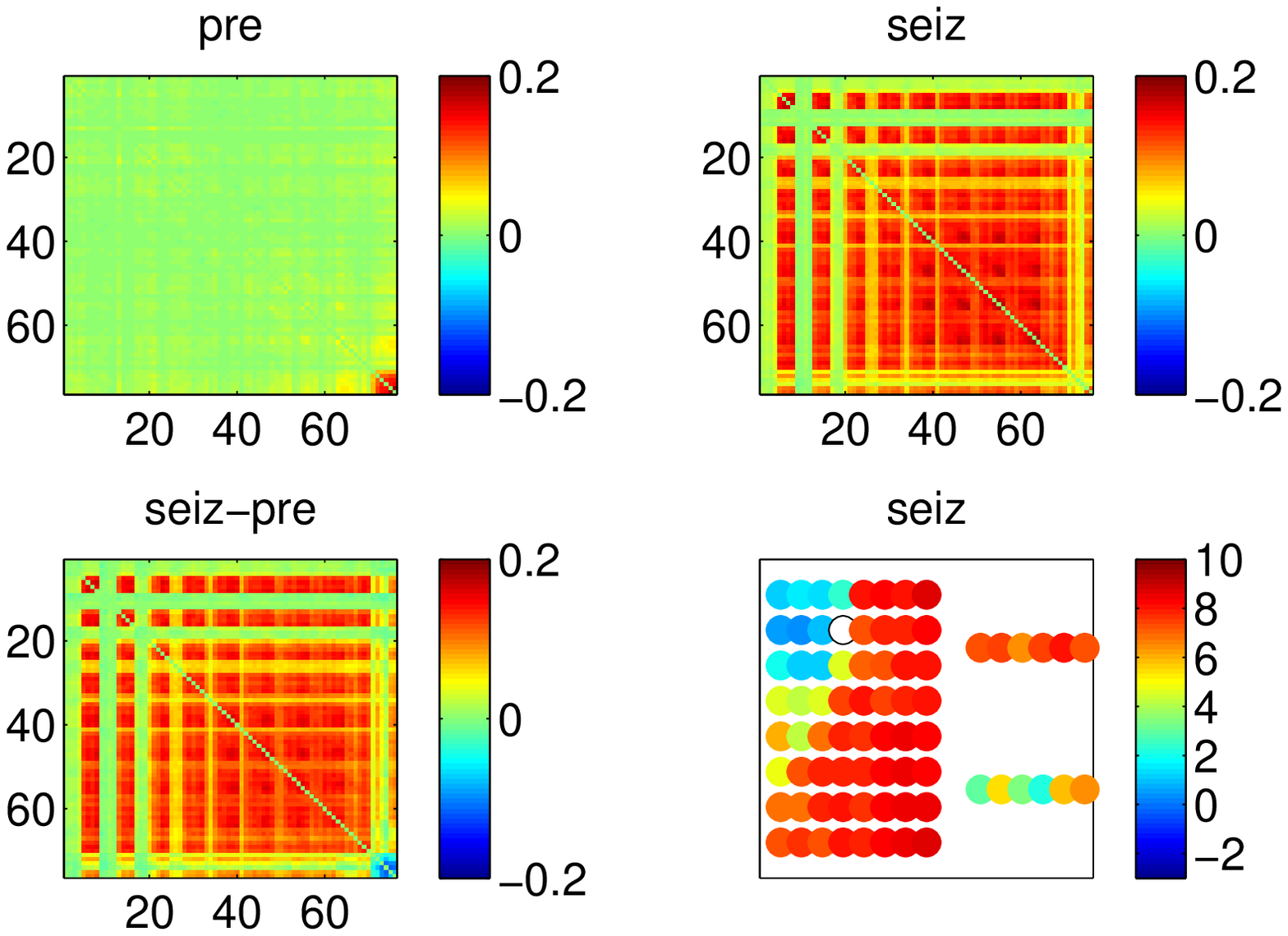}
\caption{{\rm The lagged second order terms $B^0_{ij}$ for a
cortical electrode (in white) right before and during the clinical
onset of a seizure, and the sum over the second electrode of the
pair in the lower right panel. \label{figcort}}}\end{figure}

\begin{figure}[ht]
\includegraphics[width=8.5cm]{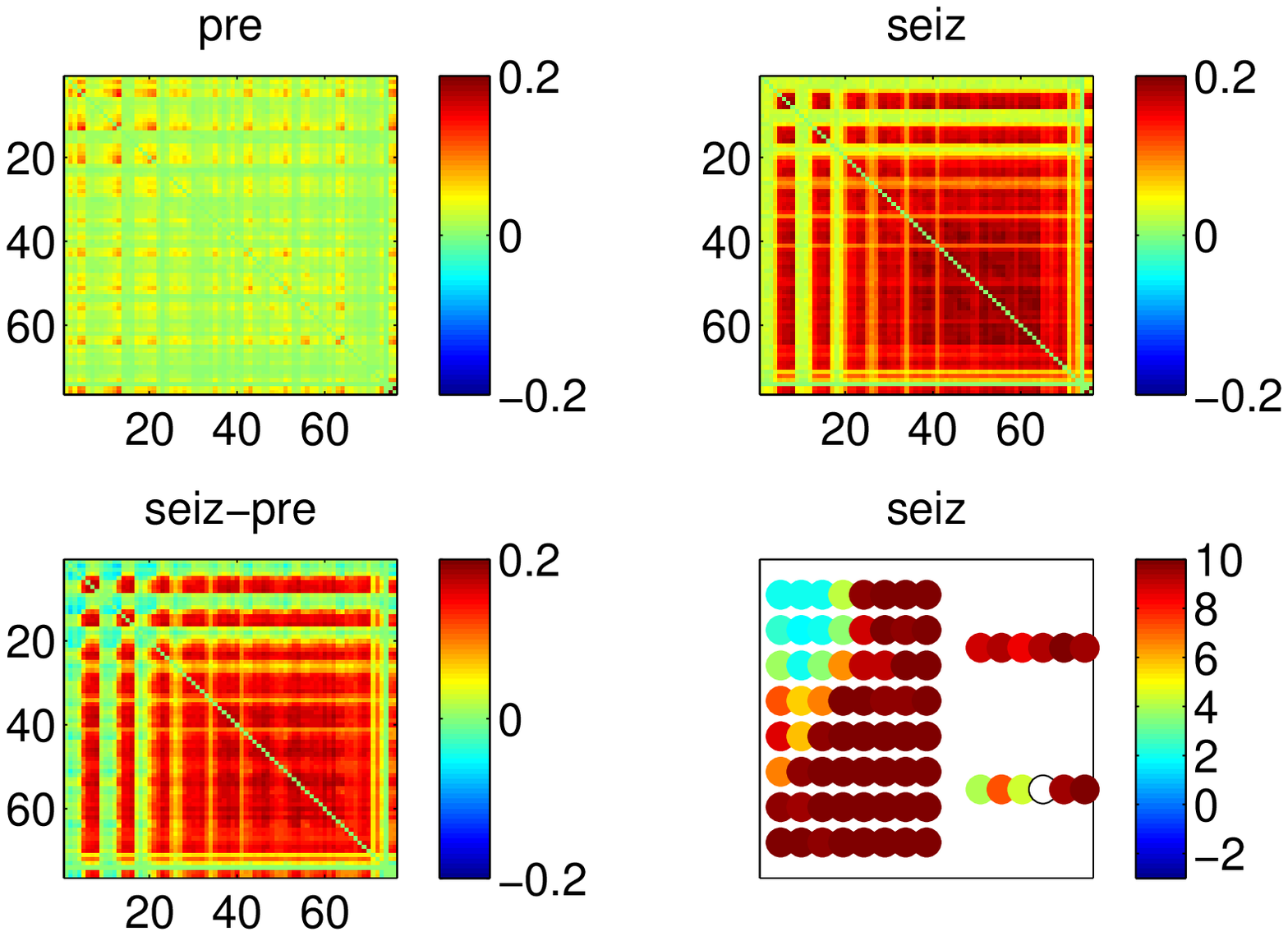}
\caption{{\rm The lagged second order terms $B^0_{ij}$ for a depth
electrode (in white) right before and during the clinical onset of a
seizure, and the sum over the second electrode of the pair in the
lower right panel. \label{figdepth}}}\end{figure}

\begin{figure}[ht]
\includegraphics[width=8.5cm]{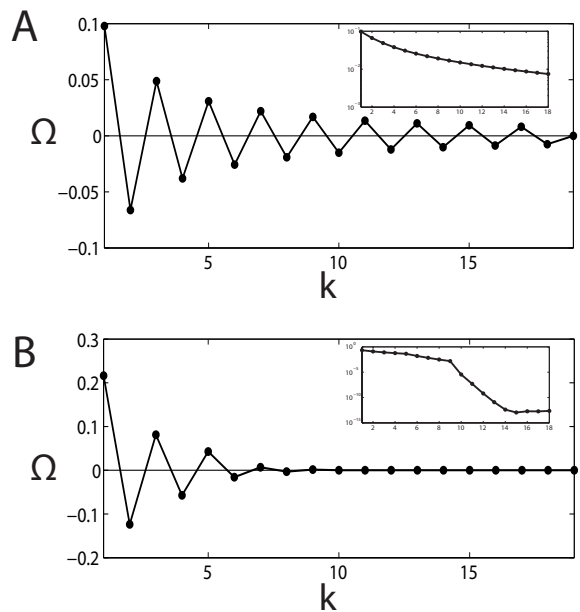}
\caption{{\rm $\Omega_k$ as a function of the multiplet size $k$ for
a model in which one variable is influenced by all the other
variables or by part of them. In (A) $m=20$ and $M=0$: all the 20
variables influence the target with unitary weight. In (B) $m=10$
and $M=10$; the weights $b_\alpha$ are [ 1.75 1.75 1 1 1 1 .5 .5 .5
.5 ]. The insets show the logarithm of the absolute value of
$\Omega_k$. The first point $k=1$, in both plots, represents the
initial pair of variables chosen as the seed, i.e. $\{1, 2\}$. The
other parameters are, in both cases, $a, \sigma,\sigma_1,\sigma_2
=0.5$. \label{figmodels}}}\end{figure}

\begin{figure}[ht]
\includegraphics[width=8.5cm]{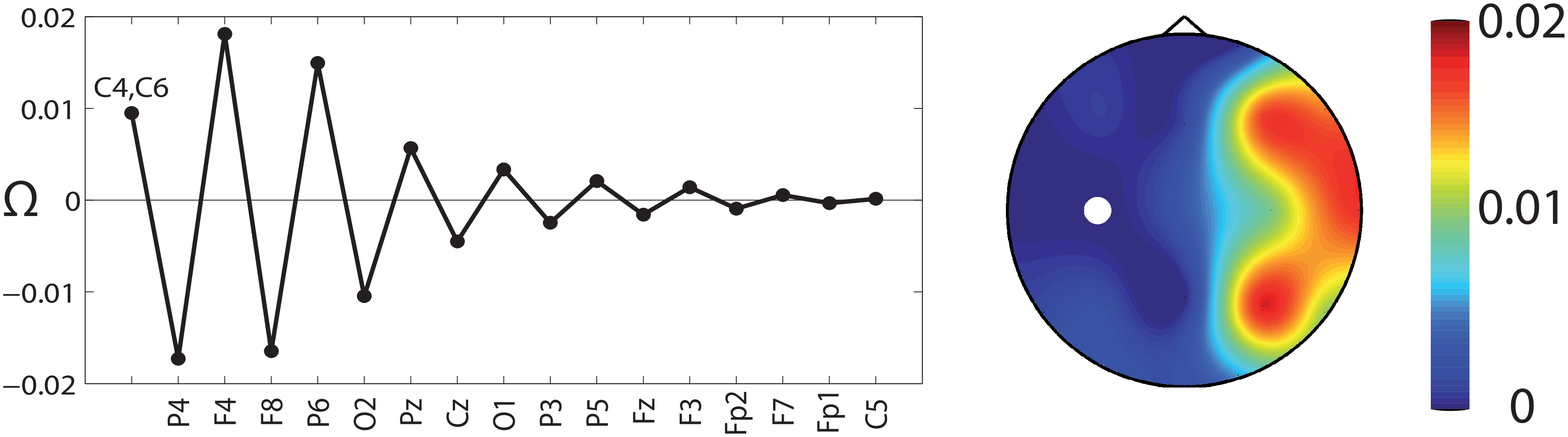}
\caption{{\rm Informative contributions to the target electrode C3.
Left: information contribute from the resulting multiplet when time
series from a given electrode are added to the existing multiplet,
starting from the pair (C4,C6) which is the one which shares the
most of information on the future of the target time series.
Channels P4, F4, F8, P6, O2, Pz and Cz are recognized to belong to
the same multiplet as C4 and C6, whilst including $O1$ leads to a
$\Omega_k$ which is not significantly different from zero.  Right:
the absolute value of this contributions plotted on a scalp map.
 \label{figC3}}}\end{figure}

\begin{figure}[ht]
\includegraphics[width=8.5cm]{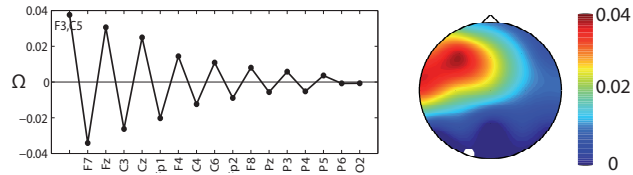}
\caption{{\rm Informative contributions to the target electrode O1.
Left: information contribute from the resulting multiplet when time
series from a given electrode are added to the existing multiplet,
starting from the pair (F3,C5) which is the one which shares the
most of information on the future of the target time series. Nine
channels (F7, Fz, C3, Cz, Fp1, F4, C4, C6, Fp2) are recognized to
belong to the same multiplet, for the remaining variables $\Omega_k$
is not significantly different from zero. Right: the absolute value
of this contributions plotted on a scalp map.
 \label{figO1}}}\end{figure}

\section{CONCLUSIONS}
Summarizing, we have proposed to describe the flow of information,
in a system, by means of multiplets of variables which  send
information to each assigned target node. We used a recently
proposed expansion of the mutual information, between a stochastic
variable and a set of other variables, to measure the character and
the strength of  multiplets of variables. Indeed, terms of the
proposed expansion put in evidence irreducible sets of variables
which provide information for the future state of the target
channel. The sign of the contributions are related to their
informational character; for the second order terms, synergy and
redundancy correspond to negative and positive sign, respectively.
For higher orders, we have shown that groups of variables, related
to the same source of information, lead to contributions with
alternating signs as the number of variables is increased.  A
decomposition with similarities to the present work have been
reported in \cite{beer}, where for multiple sources the distinction
between unique, redundant, and synergistic transfer has been
proposed; in \cite{lizier} the inference of an effective network
structure, given a multivariate time series, using incrementally
conditioned transfer entropy measurements, has been discussed.
The main purpose of this paper is to introduce an information based decomposition, and we did that in a framework unifying Granger causality and Transfer entropy, thus using a formula which is exact for linear models. In cases in which a nonlinear model is required, the entropy has to be computed, requiring a high enough number of time points for statistical validation; nonetheless the expansion that we proposed remains valid and exact in both cases.

We have reported the results of the applications to two EEG
examples. The first data set is from {\it resting brains} and we
found signatures of inter-hemispherical communications and frontal
to occipital flow of information. Concerning a data set from an
epileptic subject, our analysis puts in evidence that the seizure is
already active, close to the primary lesion, before it is clinically
observable.


\begin{thebibliography}{99}
\bibitem{te} T. Schreiber,{\it Phys. Rev.
Lett.} {\bf 85}, pp. 461-464, 2000.
\bibitem{granger} C.W.J. Granger, {\it Econometrica} {\bf 37}, pp. 424-438, 1969.
\bibitem{lehnertz} M. Staniek, K. Lehnertz, Phys. Rev. Lett. {\bf 100},
158101 (2008).
\bibitem{wibral} M. Lindner, R. Vicente, V. Priesemann, and M.
Wibral, BMC Neuroscience {\bf 12}, 119 (2011).
\bibitem{blinoska} K.J. Blinowska, R. Kus, M. Kaminski,  {\it Phys. Rev. E} {\bf 70},
pp. 50902-50905(R), 2004.
\bibitem{smirnov} D.A. Smirnov, B.P. Bezruchko,  {\it Phys Rev.} {\bf E 79}, pp. 46204-46212,
2009.
\bibitem{dingprl} M. Dhamala, G. Rangarajan, M. Ding,{\it Phys. Rev. Lett.} {\bf 100}, pp. 18701-18704, 2008.
\bibitem{noiprl} D. Marinazzo, M. Pellicoro, S. Stramaglia, {\it Phys. Rev. Lett.} {\bf 100}, pp. 144103-144107,
2008.
\bibitem{faes}L. Faes, A. Porta, G. Nollo, {\it Phys. Rev.} {\bf E 78}, 26201 (2008).
\bibitem{porta}A. Porta et al,  {\it  Methods of
Information in Medicine} 49, pp. 506-510, 2010.
\bibitem{barnett} L. Barnett, A.B. Barrett, and A.K. Seth, {\it
Phys. Rev. Lett.} {\bf 103}, pp. 238701-23704, 2009.
\bibitem{hla}K. Hlavackova-Schindler, M. Palus, M. Vejmelka,
J. Bhattacharya, Physics Reports {\bf 441}, 1 (2007).
\bibitem{noired} L. Angelini, M. de Tommaso, D. Marinazzo, L. Nitti,
M. Pellicoro, and S. Stramaglia, {\it Phys. Rev.} {\bf E 81}, 37201
(2010).
\bibitem{borst}
A. Borst, F.E. Theunissen, {\it Nat. Neurosci.} {\bf 2}, pp.
947-957, 1999.
\bibitem{sch} E. Schneidman, W. Bialek, M.J. Berry II,  {\it J.
Neuroscience} {\bf 23}, pp. 11539-11553, 2003.
\bibitem{bett} L.M.A. Bettencourt et al.,  {\it Phys.
Rev.} {\bf E 75}, pp. 21915-21924, 2007.
\bibitem{milo}R. Milo et al.,  {\it Science} 298, pp. 824-827,
2002.
\bibitem{yeger}E. Yeger-Lotem et al., {\it Proc. Natl.Acad. Sci.
U.S.A.} 101, pp. 5934-5939, 2004.
\bibitem{bettencourt} L.M.A. Bettencourt, V. Gintautas, M.I. Ham, {\it Phys. Rev. Lett.} {\bf 100}, pp. 238701-238704, 2008.

\bibitem{mcgill} W.J. McGill, ``Multivariate information transmission" {\it Psychometrika} {\bf 19}, 97-116 (1954).

\bibitem{bell} A.J. Bell , ``The co-information lattice". Proceedings of ICA 2003, Nara, Japan, April 2003.

\bibitem{shehzad2009} Z. Shehzad et al., {\it Cerebral cortex} {\bf 19}, pp. 2209-2229,
2009.
\bibitem{Tzourio2002} N. Tzourio-Mazoyer et al.,  {it Neuroimage} {\bf 15}, pp. 273-289,
2002.
\bibitem{Fox2005} M.D. Fox et al., {\it Proc Natl Acad Sci U S A.} {\bf 102}, pp. 9673-9678, 2005.
\bibitem{Salvador2005} R. Salvador et al.,  {\it Cereb. Cortex} {\bf 15}, 1332-1342,
2005.
\bibitem{target} R. Leech, R. Braga, and D. J. Sharp,  {\it The Journal of Neuroscience}  32, pp. 215-222, 2012.
\bibitem{nolte} G. Nolte et al.,
 {\it  Phys. Rev. Lett.} {\bf 100}, pp.
234101-234104, 2008.
\bibitem{website_nolte} http://clopinet.com/causality/data/nolte/
\bibitem{nolte} G. Nolte, A. Ziehe, V. Nikulin, A. Schl\"{o}gl, N. Kr\"{a}mer, T. Brismar, K.R. M\"{u}ller,
Phys. Rev. Lett. {\bf 100}, 234101, 2008

\bibitem{website_nolte} http://clopinet.com/causality/data/nolte/, accessed may 2012

\bibitem{kol_paper} M.A. Kramer, E.D. Kolaczyk, H.E. Kirsch, Epilepsy Research {\bf 79}, 173, 2008

\bibitem{kol_web} http://math.bu.edu/people/kolaczyk/datasets.html, accessed may
2012

\bibitem{beer} P.L. Williams, R.D. Beer, {\it Generalized Measures of Information
Transfer}, preprint arXiv:1102.1507 (2011).

\bibitem{lizier} J. T. Lizier, M. Rubinov, "Multivariate construction of
effective computational networks from observational data", preprint
MPI MIS Preprint 25/2012.

\end{thebibliography}
\end{document}